\documentclass[aps,prx,floatfix,superscriptaddress,noshowpacs,twocolumn]{revtex4-2}

\usepackage{graphics,graphicx,epsfig,amsmath,amssymb,epstopdf,wasysym,comment}
\usepackage{hyperref}
\usepackage{color}
\usepackage[normalem]{ulem}
\usepackage[dvipsnames]{xcolor}
\usepackage{braket,amsfonts}
\usepackage{dsfont}
\usepackage{hyperref}
\usepackage[dvipsnames]{xcolor}
\usepackage{amsthm}
\theoremstyle{remark}

\allowdisplaybreaks

\newcommand{\be}{\begin{equation}}
\newcommand{\ee}{\end{equation}}
\newcommand{\bea}{\begin{eqnarray}}
\newcommand{\eea}{\end{eqnarray}}
\newcommand{\ba}{\begin{eqnarray*}}
\newcommand{\ea}{\end{eqnarray*}}
\newcommand{\dagga}{{\phantom{\dagger}}}
\newcommand{\bR}{\mathbf{R}}

\newcommand{\bq}{\mathbf{q}}

\newcommand{\bk}{\mathbf{k}}

\newcommand{\bp}{\mathbf{p}}

\newcommand{\bRp}{\mathbf{R'}}

\newcommand{\br}{\mathbf{r}}
\newcommand{\Ima}{\text{Im}}

\newcommand{\dis}{\displaystyle}

\newcommand{\up}{\uparrow}
\newcommand{\down}{\downarrow}
\newcommand{\fract}[2]{\frac{\dis #1}{\dis #2}}

\newcommand{\eqn}[1]{(\ref{#1})}

\renewcommand{\bra}[1]{\langle #1 \mid}
\renewcommand{\ket}[1]{\mid #1 \rangle}
\newenvironment{eqs}%
{\begin{equation} \begin{aligned}}%
{\end{aligned} \end{equation} }
\newcommand{\beal}{\begin{eqs}}
\newcommand{\eal}{\end{eqs}}
\newcommand{\bealn}{\beal\nonumber}
\newcommand{\bw}{\begin{widetext}}
\newcommand{\ew}{\end{widetext}}
\newcommand{\esp}[1]{\text{e}^{#1}}

\newcommand{\ep}{\epsilon}
\newcommand{\bd}[1]{{\boldsymbol{#1}}}
\newcommand{\bkappa}{\bd{\kappa}}

\begin{document}
\title{Electrical Transport in the Hatsugai-Kohmoto Model}

\author{Daniele Guerci}
\affiliation{Center for Computational Quantum Physics, Flatiron Institute, New York, NY 10010}
\author{Giorgio Sangiovanni}
\affiliation{Institut f\"ur Theoretische Physik und Astrophysik, Universit\"at W\"urzburg, 97074 W\"urzburg, Germany} 
\author{Andrew J. Millis}
\affiliation{Department of Physics, Columbia University, New York, NY 10027}
\affiliation{Center for Computational Quantum Physics, Flatiron Institute, New York, NY 10010}
\author{Michele Fabrizio}
\affiliation{International School for Advanced Studies (SISSA), Via Bonomea 265, I-34136 Trieste, Italy} 

\begin{abstract}
We show that in models with the Hatsugai-Kohmoto type of interaction that is local in momentum space thus infinite-range in real space, Kubo formulas neither reproduce the correct thermodynamic susceptibilities, nor yield sensible transport coefficients. Using Kohn's trick to differentiate between metals and insulators 
by threading a flux in a torus geometry, we uncover the striking property that Hatsugai-Kohmoto models with an interaction-induced gap in the spectrum sustain a current that grows as the linear size at any non-zero flux and which can be either diamagnetic or paramagnetic.  
\end{abstract}

                         
\maketitle

\section{Introduction}
In periodic systems, many-body electronic correlations induce insulating phases that can either be of genuine Mott character ~\cite{Mott-1949} or accompanied by translational and/or time reversal symmetry breaking~\cite{Wigner-1934}. The underlying mechanism is the non-perturbative interplay between band energy gain, which  
is maximised by electron delocalisation throughout the lattice, and Coulomb repulsion, which instead is minimised by electron localisation.
While the former prefers many-body wavefunctions 
factorised in crystalline momentum $\bk$, the latter favours instead factorisation in real space.
This competition escapes conventional independent-particle descriptions, making any theoretical approach to the interaction-driven transition from free electrons to Mott- or Wigner-like insulators intrinsically  difficult. \\
As a consequence, the search for minimal models capturing key physical aspects of this transition has been considered essential ever since the earliest theoretical as well as experimental instances of Mottness. In this respect, the chief role has been for long played by the single-band Hubbard model~\cite{Hubbard-1963}, where the band energy is provided by inter-site hopping and the Coulomb repulsion replaced by an on-site one. This model, a lynchpin of the research on strongly correlated materials, eludes an exact solution except in one or infinite spatial dimensions despite its simplicity.
Hatsugai and Kohmoto~\cite{HK_original} introduced a variant of the Hubbard model which is exactly solvable and displays an interaction-driven gap-opening transition; this model has attracted considerable further attention in the context of the Mott transition~\cite{HK_original,CONTINENTINO1994619,PhysRevB.61.7941,PhysRevD.99.094030} and, more recently, superconductivity \cite{Tarasewicz,P2_supercond,Li_2022} and topological band structure effects \cite{PhysRevResearch.5.013162,PhysRevB.108.035121,Phillips-PRL2023,Jan-PRB2023,Qimiao-preprint2023,Jan-PRB2024}. 
Variations starting from the HK but introducing a coupling between high-symmetry momenta have been also proposed for the analysis of the Luttinger surface \cite{wormHK}. 
Models with HK interactions allow analytically access not only to the thermodynamic properties of Mott phases~\cite{HK_original,CONTINENTINO1994619,PhysRevB.61.7941,PhysRevD.99.094030}, but also their single-particle Green's functions $G(\ep,\bk)$, which in the gapped phase exhibit bands of Green function zeros~\cite{guraire2011,volovik2012topologyquantumvacuum,fabrizio2023,wagner2023_ncomm,wagner2023edgezerosboundaryspinons} a hard-to-compute non-perturbative effect that can remarkably be obtained by means of simple exact calculations with the HK interaction. 
The peculiar feature of the HK interaction that makes the model integrable in any dimension and for generic number of bands is its locality in momentum space, $U_\text{HK}=\sum_{\bk}U_\text{HK}(\bk)$, which corresponds to infinite range in real space~\cite{PhysRevB.101.184506}. 
Such a decisive modification of the spatial structure of the electron-electron repulsion calls for a closer examination of some physical properties of HK interacting models. 



In this paper we focus on the fundamental case of long wavelength density-density and current-current response functions, exposing a fundamental non-physical aspect of models with HK type interactions and showing that the Mott phase of HK models in fact has a divergent long wavelength current response physically related to an infinite-ranged pair hopping implied by the HK form of the interaction and mathematically expressed by a nonanaliticity in the $\bq\to0$ limit of response functions. The rest of this paper is organized as follows: in Section~\ref{sec1} we present general arguments highlighting the importance of the analyticity of the long-wavelength limit in the extraction of physical observables from Kubo-type linear response formulas. Section~\ref{sec2} defines the generalized HK models that we analyse and recalls their basic properties. Section~\ref{sec3} applies to HK models a criterion introduced by Kohn in 1964~\cite{Kohn-1964} to distinguish metals from insulators, showing that the gapped phases of HK models are in fact metallic, with a non-vanishing Drude weight. Section~\ref{sec4} is a summary and conclusion.

\section{Charge conservation versus continuity equation}\label{sec1}

We begin with some general remarks relating to the long wavelength limit of correlation functions of conserved quantities. We explicitly discuss quantities related to the particle density but the results can be straightforwardly generalized to any other conserved quantities, e.g. the total energy or (for models without spin-orbit interaction), the total spin. 

Assume a many-body Hamiltonian $H$ that commutes with the total number of electrons $N= \int d\br\, \rho(\br)$ with $\rho(\br)$ the particle density operator. We define the Fourier transform of $\rho(\br)$ as $\rho(\bq)= \int d\br\, \esp{-i\bq\cdot\br}\;\rho(\br),$
thus
\beal
\Big[\rho(\bq=\bd{0})\,,\,H\Big] &=\Big[N\,,\,H\Big] = 0\,.
\label{q=0}
\eal
If, as is in the case of conventional models with finite range Coulombic interactions, $\bq\to0$ limit is smooth,
 \beal
\lim_{\bq\to \bd{0}}\,\Big[\rho(\bq)\,,\,H\Big] &=0\,,
\label{q to 0}
\eal
then many important properties follow. For example, the static structure factor at zero temperature, which we define hereafter as   
\be
S(\bq) = \Big(\bra{\Psi_0} \rho(\bq)\,\rho(-\bq)\ket{\Psi_0} - N^2\,\delta_{\bq\bd{0}}\Big),
\label{S(q)}
\ee
with $\ket{\Psi_0}$ the ground state, obeys
\be
\lim_{\bq\to0}S(\bq)=0.
\ee
Similarly, the retarded density-density response function at zero temperature is defined as 
\be
\chi(\omega,\bq) = -\fract{i}{V}\int_0^\infty \!\!\!\!dt \,\esp{i\omega t}\,
\bra{\Psi_0\!} \!\Big[\rho(t,\bq),\rho(-\bq)\Big]\!\ket{\!\Psi_0}\,,
\label{chi}
\ee
with $V$ the number of $\bk$ points within the first Brillouin zone, and $\omega$ having an infinitesimal positive imaginary part. Smoothness in $\bq$ implies that 
\beal
\chi(\omega,\bq\to\bd{0}) &= 0\,,& \forall&\,\omega\not=0\,,
\label{chi-q=0}
\eal
as well as 
\beal
\fract{\partial n}{\partial \mu} &=-\lim_{\bq\to\bd{0}}\,\chi(0,\bq)\,,
\label{compressibility}
\eal
where $n=N/V$ and $\partial n/\partial\mu$ is the thermodynamic charge compressibility.
One particularly important consequence of the smoothness of the $\bq\to0$ limit is
\beal
\Big[\rho(\bq)\,,\,H\Big] 
\equiv \bq\cdot\bd{J}(\bq)\,,
\label{continuity-q}
\eal
with $\bd{J}(\bq\to \bd{0})$ non singular. 
Eq.~\eqn{continuity-q} is just the continuity equation, which allows the identification of the longitudinal component of the current operator $\bd J(\bq)$.
The smoothness of the $\bq\to0$ limit ensures that at $\bq=0$ the current operator defined in this way is identical to the current operator defined from $\delta H/\delta \bk$ with the minimal coupling $\bk\to\bk-e\bd A/c$ and $\bd A$ the vector potential. 
 
To summarize this section, one can define a meaningful current operator if any of equations \eqn{S(q)}, \eqn{chi-q=0} and \eqn{compressibility} is satisfied. If one insists in using \eqn{continuity-q} to find 
$\bd{J}(\bq)$ even when they are violated, the outcome is a current operator that is singular for $\bq\to\bd{0}$, whose physical consequences we here uncover 
in models with HK interactions.

\vspace{0.5cm}

\section{Models with Hatsugai-Kohmoto type of interactions}\label{sec2}

\begin{figure}
\centering 
\includegraphics[width=0.9\linewidth]{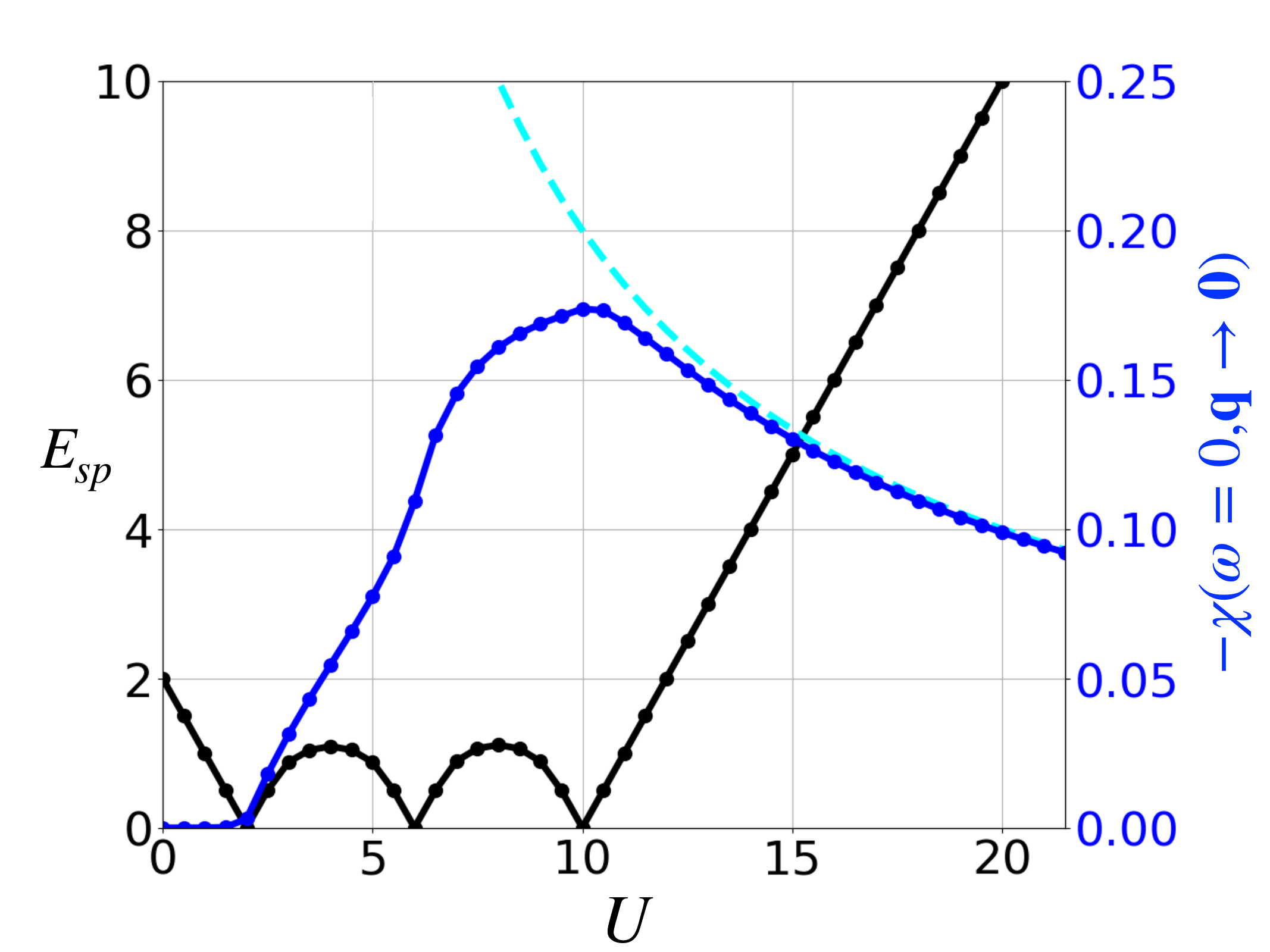}
\caption{Single-particle energy gap $E_{sp}$ (left axis) and static limit of the density-density response function,  
$-\chi(\omega=0,\bq\to\bd{0})$, (right axis), for the HK interacting BHZ model introduced in the text, as function of $U$ at $t=1$, the energy unit, $t'=0.4$, $M=3$ and $J=0.001$.  The dashed cyan line shows the large $U$ result $-\chi(\omega=0,\bq\to0)\approx 1/(U+3J/2)$.}
\label{fig:chi_dd}
\end{figure}

We consider $M$ (counting both spin and orbital degrees of freedom) bands of interacting electrons with Hamiltonian 
\beal
H &= H_0+U_\text{HK} =\sum_\bk\,H(\bk)\\
&=\sum_\bk\,\Big( H_0(\bk) + U_\text{HK}(\bk)\Big)\,,
\label{Ham}
\eal
where 
\be
H_0(\bk) =\sum_{a,b=1\dots M} t^{ab}(\bk)\, c^\dagger_{a\bk}\, c^\dagga_{b\bk}\,,
\label{Ham-0}
\ee
is the non-interacting part, whereas 
\be
U_{\rm HK}(\bk) =\frac{1}{2}\sum_{a,b,c,d} U_{abcd} ({\bk}) \,c^\dagger_{a\bk}\, c^\dagger_{b\bk}\,c^\dagga_{c\bk}\, c^\dagga_{d\bk}\,,
\ee
is a generic HK two-body interaction~\cite{HK_original}, local in momentum space. Here $c^\dagger_{a\bk}$ creates an electron into a state with wave function $\phi_{a\bk}(\br)$ and for later reference the density operator 
\be
\rho(\bq)=\sum_{\bk a b}\,c^\dagger_{a\bk}\,\rho_{a\bk,b\bk+\bq}\,
c^\dagga_{b\bk+\bq},
\ee
with coefficients 
\be
\rho_{a\bk,b\bk+\bq}=\int d\br\, \phi_{a\bk}(\br)^*\,\esp{-i\bq\cdot\br}\;\phi_{b\bk+\bq}(\br)\underset{\bq\to 0}{\to}
\delta_{ab},\label{coefficients rho}
\ee
where the second line follows from the orthogonality of wave functions. 
We observe that, in real space $U_\text{HK}$ corresponds to
\bw
\beal
&U_\text{HK}=\sum_{abcd}\fract{1}{V}\,\int d\bR\,d\bR'\,d\br\,d\br'\,
U_{abcd}(\bR-\bR')\;
\Psi_a^\dagger\left(\frac{\bR}{2}+\br\right)\,\Psi_b^\dagger\left(\frac{\bR}{2}-\br\right)\,
\Psi^\dagga_c\left(\frac{\bRp}{2}-\br'\right)\,\Psi^\dagga_d\left(\frac{\bRp}{2}+\br'\right)\,,
\label{HK-interaction}
\eal
\ew
where $\Psi^\dagger_a(\br) = \sum_\bk\,\phi_{a\bk}(\br)\,c^\dagger_{a\bk}$ is the operator that creates an orbital-$a$ electron at position $\br$. We thus see that the HK interaction includes pair hopping and orbital exchange terms with peculiar long-ranged structure. A weak $\bk$ dependence is normally assumed, so that the center of mass of a pair is preserved, $\bR\simeq \bRp$, still the interaction is independent of the relative spacing of particles in a pair thus can move charge over arbitrarily large distances, for example annihilating a close pair (small relative separation $\br'$) and creating a widely separated pair 
(large $\br$). \\
Because $H(\bk)$ conserves the number of electrons in each $\bk$ sector, we may label the eigenstates by crystal momentum $\bk$, $n_{\bk}=1,\dots,M$ the number of particles. Within the sector with $n_{\bk}$ particles, we denote as $\ket{\!\ell_{\bk},\bk,n_{\bk}}$ the eigenstates of $H(\bk)$ with eigenvalues $E_{\ell_{\bk}}(\bk,n_{\bk})$ in ascending order in $\ell_{\bk}=0,\dots,\binom{2M}{m}-1$. A generic many-body eigenstate of $H$ is therefore a product 
state
\bealn
\ket{\{\ell_\bk,n_\bk\}}\equiv \prod_\bk\,\ket{\ell_\bk,\bk,n_\bk}\,,
\eal
with energy $\sum_\bk\,E_{\ell_\bk}(\bk,n_\bk)$ and number of electrons 
$\sum_\bk\,n_\bk$.  \\
Let us label the (possibly degenerate) ground state as
\bealn
\ket{\Psi_0} &\equiv \prod_\bk\,\ket{\ell_{0\bk},\bk,n_{\bk}}\,,
\eal
where at each $\bk$ the occupancy $n_{\bk}$ and the internal quantum number $\ell_{0\bk}$ are chosen to minimize $H(\bk)$. 
It is very useful for subsequent discussion to define the natural orbitals $\alpha$ with corresponding creation $c^\dagger_{\alpha\bk}$ and annihilation $c_{\alpha\bk}$ operators where $\alpha=1,\dots,M$ in such a way that
\beal
\bra{0,\bk,n} c^\dagger_{\alpha\bk}\,c^\dagga_{\beta\bk}\ket{0,\bk,n}
&= \delta_{\alpha\beta}\,n_\alpha(\bk)\,,
\label{natural basis}
\eal
with $\sum_\alpha\,n_\alpha(\bk)=n$, and assuming $n<M$. Correspondingly, we define 
\beal
\bra{\ell,\bk,n-1}c^\dagga_{\alpha\bk}\ket{0,\bk,n}&= 
\sqrt{n_\alpha(\bk)\;}\;C^-_{\ell\alpha}(\bk)\,,\\
\bra{\ell,\bk,n+1}c^\dagger_{\alpha\bk}\ket{0,\bk,n}&= 
\sqrt{1-n_\alpha(\bk)\;}\;C^+_{\ell\alpha}(\bk)\,,
\label{def}
\eal
where, because of \eqn{natural basis}, 
\be
\sum_\ell\,C^+_{\ell\alpha}(\bk)^*\,C^+_{\ell\beta}(\bk) = \sum_\ell\,C^-_{\ell\alpha}(\bk)^*\,C^-_{\ell\beta}(\bk)=\delta_{\alpha\beta}\,,
\label{norm}
\ee
and thus the coefficients $C^\pm_{\ell\alpha}(\bk)$ are, by construction, of order one. In terms of the natural orbitals creation and annihilation operators, the 
Fourier transform of the density reads
\beal
\rho(\bq) &= \sum_{\bk\alpha\beta}\,c^\dagger_{\alpha\bk}\;
\rho_{\alpha\bk,\beta\bk+\bq}\;
c^\dagga_{\beta\bk+\bq}\,,
\label{rho(q)}
\eal
where $\rho_{\alpha\bk,\beta\bk+\bq}$ is defined as in \eqn{coefficients rho}.
The static structure factor \eqn{S(q)} for $\bq\not=\bd{0}$ is therefore
\bealn
S(\bq) &= \bra{\Psi_0}\rho(\bq)\,\rho(-\bq)\ket{\Psi_0} \\
&= \sum_{\bk\bp}\,\sum_{\alpha\beta\gamma\delta}\,
\rho_{\alpha\bk,\beta\bk+\bq}\,
\rho_{\delta\bp,\gamma\bp+\bq}^*\\
&\qquad \qquad
\bra{\Psi_0} c^\dagger_{\alpha\bk}\,c^\dagga_{\beta\bk+\bq}\;
c^\dagger_{\gamma\bp+\bq}\,c^\dagga_{\delta\bp}\ket{\Psi_0}\,.
\eal
Since the ground state is factorised, we readily find through \eqn{natural basis} and \eqn{coefficients rho} that 
\bealn
S(\bq) &= \sum_{\bk\alpha\beta}\, \big|\rho_{\alpha\bk,\beta\bk+\bq}\big|^2\,
n_\alpha(\bk)\,\big(1-n_\beta(\bk+\bq)\big)\\
&\xrightarrow[\bq\to\bd{0}]{}\,\sum_{\bk\alpha}\, n_\alpha(\bk)\,\big(1-n_\alpha(\bk)\big)
\,,
\eal
hence does not vanish for $\bq\to\bd{0}$ unless $n_\alpha(\bk)=0,1$, 
which occurs, e.g., if the model is non interacting, $H^\text{HK}_{int}(\bk)=0$ (recall that we assume zero temperature). \\
The density-density response function \eqn{chi}, using \eqn{def} 
and defining the matrix elements
\beal
\text{P}_{\ell\bk+\bq,\ell'\bk} 
&= \sum_{\alpha\beta}\,\rho_{\beta\bk,\alpha\bk+\bq}^*\,
C^+_{\ell\alpha}(\bk+\bq)\,C^-_{\ell'\beta}(\bk)\\
&\qquad\qquad \sqrt{\big(1-n_\alpha(\bk+\bq)\big)\,n_\beta(\bk)\;}\;,
\label{P}
\eal
reads 
\beal
\chi(\omega,\bq) &= \fract{1}{V}\!\sum_{\bk,\ell\ell'}\!
\Bigg\{
\fract{\;\text{P}^*_{\ell\bk+\bq,\ell'\bk}\,\text{P}^\dagga_{\ell\bk+\bq,\ell'\bk}\;}{\omega-\ep_\ell(\bk+\bq,n+1)-\ep_{\ell'}(\bk,n-1) }\\
&\qquad \qquad 
- \fract{\;\text{P}^*_{\ell\bk,\ell'\bk+\bq}\,\text{P}^\dagga_{\ell\bk,\ell'\bk+\bq}\;}{\omega+\ep_\ell(\bk,n+1)+\ep_{\ell'}(\bk+\bq,n-1) }
\Bigg\}\,,
\label{chi-HK}
\eal
where 
\bealn
\ep_\ell(\bk,n\pm1)=E_\ell(\bk,n\pm 1) -E_0(\bk,n)>0\,,
\eal
are the excitation energies. For $\bq\to\bd{0}$, $\chi(\omega,\bq)$ in \eqn{chi-HK} 
becomes 
\bealn
\chi(\omega,\bq\to\bd{0}) &\to \fract{1}{V}\sum_{\bk,\ell\ell'}\,
\Bigg\{ \;\big|\text{P}^\dagga_{\ell\bk,\ell'\bk}\big|^2
\\&\qquad 
\fract{\ep_\ell(\bk,n+1)+\ep_{\ell'}(\bk,n-1)}
{\;\omega^2-\big(\ep_\ell(\bk,n+1)+\ep_{\ell'}(\bk,n-1)\big)^2\;}\Bigg\}\;,
\eal
with  
\beal
\text{P}_{\ell\bk,\ell'\bk} 
&= \sum_{\alpha}\,
C^+_{\ell\alpha}(\bk)\,C^-_{\ell'\alpha}(\bk)\\
&\qquad\qquad \sqrt{\big(1-n_\alpha(\bk)\big)\,n_\alpha(\bk)\;}\;.
\label{rho-matrix-elements}
\eal
Therefore, $\chi(\omega,\bq\to\bd{0})$ is evidently non zero if $n_\alpha(\bk)\not=0,1$, see \eqn{rho-matrix-elements}, in contrast to \eqn{chi-q=0}. 
Suppose, as we do hereafter, that the model describes an insulator with a single-particle gap, thus vanishing compressibility $\partial n/\partial\mu$ and 
$\ep_\ell(\bk,n\pm 1)>0$ that remain finite in the thermodynamic limit.
However, looking at Eq.~\eqn{chi-HK} we readily conclude that 
\bealn
0 = \fract{\partial n}{\partial\mu} \not= -\lim_{\bq\to\bd{0}}\,\chi(0,\bq)\,.
\eal
It follows that, in presence of interaction, $\chi(0,\bq\to\bd{0})$ does not provide the correct charge compressibility.\\
This is directly tested on a two-orbital Hamiltonian 
$H=\sum_{\bk} \big(H_0(\bk)+U_{\rm HK}(\bk) \big)$, 
where $H_0(\bk)$ is a non-interacting BHZ \cite{BHZ,BHZ_interacting} model, the interaction reads 
\beal
U_\text{HK}(\bk) &=\frac{U}{2}\sum_{a=1,2}\,\big(n_{a \bk }-1\big)^2+J\, {\boldsymbol S}_{\bk 1}\cdot{\boldsymbol S}_{\bk 2},
\label{two_band_model}
\eal
with $n_{\bk a}$ and $\bd{S}_{\bk a}$ the occupation number and spin operator of orbital 
$a=1,2$ at momentum $\bk$, while $J>0$ stabilises a non-magnetic Mott insulator at large $U$. 
Specifically, defining the four component spinor
$\bd{c}^\dagger_{\bk}=\big(c^\dagger_{1\bk\up},c^\dagger_{2\bk\up},
c^\dagger_{1\bk\down},c^\dagger_{2\bk\down}\big)$, 
\beal
H_0(\bk) &=  \bd{c}^\dagger_{\bk}\,
\begin{pmatrix} 
\hat{h}_\up(\bk) & 0\\
0 & \hat{h}_\down(\bk)
\end{pmatrix}\,\bd{c}^\dagga_\bk\,,
\label{BHZ_H0}
\eal
where 
\bealn
\hat{h}_{\up}(\bk)&=\big(M+t\cos k_x +t \cos k_y\big)\,\tau^z \\
& \quad +\lambda\,\big(\sin k_x \,\tau^x +\sin k_y \,\tau^y\big)\equiv{\boldsymbol d}_\up(\bk)\cdot{\boldsymbol \tau}\,,
\eal
with $\tau^i$, $i=1,2,3$, the Pauli matrices acting in the orbital space, and 
$\hat{h}_{\down}(\bk)=\hat{h}^*_{\up}(-\bk)$. At $U=0$ the model describes a quantum spin-Hall insulator, 
while $U\gg \big|{\boldsymbol d}_\sigma(\bk)\big|$ stabilises a Mott insulator that is essentially 
a collection of local-in-$\bk$ inter-orbital singlets. The Mott transition is not unique but occurs 
gradually in momentum space, starting from the $\bk$-point with the minimal non-interacting gap, until 
it affects all momenta.   
In Fig.~\ref{fig:chi_dd} we draw the single-particle gap $E_{sp}$ as function of $U$ at 
$t=1$, $\lambda=0.4$, $M=3$ and $J=0.001$. Turning on $U$, $E_{sp}$, which is finite at $U=0$, has a non-monotonous behaviour, touching zero at three values of $U$, until, above $U=10$, all 
$\bk$ points enter the Mott insulating regime and $E_{sp}$ starts growing linearly with $U$.   
In the same figure we also show the static limit of the density-density response function,  
$-\chi(\omega=0,\bq\to\bd{0})$. We observe that it coincides with the charge compressibility $\kappa$, which must vanish as long as $E_{sp}>0$, only until $E_{sp}$ reaches the first root. 
After that, the two quantities clearly deviate from each other. In particular, above $U=10$, 
the Kubo formula is finite and rapidly approaches the asymptotic behaviour $1/(U+3J/2)$.   
The above results show explicitly that any Hamiltonian with a HK type of interaction yields non analytic properties at $\bq=\bd{0}$, which, for instance, prevents from identifying a smooth current operator. 
In this respect, it is interesting to note that retardation effects in the interaction also yield inconsistencies between local and lattice Ward identities, as pointed out in Ref.~\cite{krienPRB}.

\subsection{Explicit expression of the interaction-induced current component 
and its consequences} 
To be more precise, let us write the general expression of a Hatsugai-Kohmoto type of interaction, using for convenience the operators in the natural orbital basis, 
\beal
U_\text{HK} &= \fract{1}{2}\,\sum_\bk
U_{\alpha\beta\gamma\delta}(\bk)\;c^\dagger_{\alpha\bk}\,c^\dagger_{\beta\bk}\,c^\dagga_{\gamma\bk}\,c^\dagga_{\delta\bk}\,,\label{H-int}
\eal
where repeated indices are summed and
\bealn
U_{\alpha\beta\gamma\delta}(\bk) &= -U_{\beta\alpha\gamma\delta}(\bk)
=U_{\beta\alpha\delta\gamma}(\bk)
=U_{\delta\gamma\beta\alpha}(\bk)^*\,.
\eal
Using the small $\bq$ limit of the density \eqn{rho(q)}, i.e., approximating 
the matrix element with $\delta_{\alpha\beta}$, 
one readily finds that \eqn{H-int} yields, through \eqn{continuity-q},  
a contribution $\bd{J}^\parallel_\text{int}(\bq)$ to the longitudinal component of the current operator that reads
\bea
\bd{J}^\parallel_\text{int}(\bq)
&=& \fract{\bq}{\;q^2\,}\sum_\bk\bigg\{U_{\alpha\beta\gamma\delta}(\bk+\bq)\;c^\dagger_{\alpha\bk}\,c^\dagger_{\beta\bk+\bq}\,c^\dagga_{\gamma\bk+\bq}\,c^\dagga_{\delta\bk+\bq}\nonumber\\
&&\qquad\qquad\;
-U_{\alpha\beta\gamma\delta}(\bk)\;c^\dagger_{\alpha\bk}\,c^\dagger_{\beta\bk}\,c^\dagga_{\gamma\bk}\,c^\dagga_{\delta\bk+\bq}\bigg\}\,.
\label{J-long-int}
\eea
We observe that, as discussed in detail in Ref.~\cite{Qimiao-preprint2023}, see, for comparison, Ref.~\cite{Phillips-PRL2023},
$\bd{J}^\parallel_\text{int}(\bq)$ does not vanish 
for $\bq\to\bd{0}$, as can be realised by computing the matrix elements between any two eigenstates of the Hamiltonian. 
In fact, $\bd{J}^\parallel_\text{int}(\bq)$ is singular for $\bq\to\bd{0}$, as we  anticipated.\\
This result has a striking consequence. Because of gauge invariance, 
\beal
\fract{\omega^2}{q^2}\;\chi(\omega,\bq) =
\Big(\chi_\parallel(\omega,\bq)+\text{N}(\bq)\Big)\,,
\label{gauge-invariance}
\eal
where $\chi_\parallel(\omega,\bq)$ is the longitudinal component of the 
current-current response function, and 
\bealn
\text{N}(\bq) &= \fract{1}{Vq^2}\;\bra{0}\Big[\big[\rho(\bq),H\big],\rho(-\bq)\Big]\ket{0}\,,
\eal
is the diamagnetic contribution. The left hand side of 
\eqn{gauge-invariance} vanishes for $\omega\to0$, thus 
$\chi_\parallel(0,\bq)=-\text{N}(\bq)$, which is reassuringly consistent with gauge invariance. However, because of \eqn{gauge-invariance}, the 
longitudinal component of the uniform conductivity $\sigma_\parallel(\omega)$
reads 
\bealn
\sigma_\parallel(\omega) &= \lim_{\bq\to\bd{0}}\,
\fract{ie^2}{\omega}\,\Big(\chi_\parallel(\omega,\bq)+\text{N}(\bq)\Big)\\
&=\lim_{\bq\to\bd{0}}\,
\fract{ie^2}{\omega}\;\fract{\omega^2}{q^2}\;\chi(\omega,\bq)\,.
\eal 
Hence, since $\chi(\omega,\bq\to\bd{0})\not=0$, $\sigma_\parallel(\omega)$ is singular despite the fact that the model is supposed to describe an insulator.
Such anomalous behaviour is evidently a consequence of the infinite-range Hatsugai-Kohmoto interaction and, in particular, of the long-distance pair-hopping processes that are allowed and contribute to the current. \\
The results of this and the previous section, which hold true also for densities and currents associated to other conserved quantities, show an unavoidable inconsistency in the calculation of thermodynamic susceptibilities and transport coefficients through low-frequency and long-wavelength linear response functions for models with the HK interaction. 

\section{Kohn's criterium applied to HK models}\label{sec3}
In 1964, Walter Kohn proposed a very elegant and enlightening argument \cite{Kohn-1964} to discriminate between metals and insulators through their responsiveness to boundary conditions. He argued that metals 
are sensitive to varying boundary conditions, whereas insulators are not. 
Indeed, Kohn showed \cite{Kohn-1964} that, upon twisting boundary conditions by an angle $\phi$ in 
one direction, the curvature of the total energy $E(\phi)$ at $\phi=0$ 
is proportional to the formal expression of the Drude weight $D$ of a realistic Hamiltonian, as calculated by spectral decomposition. Specifically,   
\beal
\lim_{L\to\infty}\,\fract{e^2}{L^{d-2}}\,
\fract{\partial^2 E(\phi)}{\partial\phi^2}_{\big|\phi=0} &= D \,,
\label{Kohn-Drude}
\eal
where $L$ is the system size and $d$ the dimension. Strictly speaking, for the equivalence \eqn{Kohn-Drude} to be valid, the evolution of the ground state energy with $\phi$ has to be adiabatic. The necessary but not sufficient condition for this is that the thermodynamic limit $L\to\infty$ is taken after calculating the second derivative~\cite{Scalapino-PRB1993}. Hereafter, we shall consider a system with a spectral gap to all excitations, which we believe implies that the necessary condition is also sufficient. \\ 

\noindent
We may wonder whether Kohn's distinguishing criterium holds true also in models with HK interactions, given the infinite range of the latter.  \\
For that, we consider a model on a cubic lattice with linear size $L$ and unit lattice spacing, which is folded into a three-dimensional torus. 
The lattice potential $V(\br)$ as well as the potential 
$U(\br)$ in the HK interaction \eqn{HK-interaction} are assumed periodic on the torus, e.g., $U(\br) = U(\br + L\,\bd{e}_a)$ where $\bd{e}_a$ is the unit vector in the $a=x,y,z$ direction. Therefore, also the many-body eigenfunctions of 
the Hamiltonian $H$ must have the same periodicity. \\
Next, we suppose that the torus is threaded by a flux $\phi_a$, $a=x,y,z$, which we parametrise through the vector $\bkappa = (\phi_x,\phi_y,\phi_z)/L$, so that  
the new Hamiltonian is $H(\bkappa)$. In realistic Hamiltonians 
that have, e.g., a first quantization representation, 
the flux can be formally 
gauged away by a unitary transformation $U(\bkappa)=\esp{-W(\bkappa)}$, 
$W(\bkappa)^\dagger=-W(\bkappa)$, which changes the originally periodic Fermi fields $\Psi_\alpha(\br)=\Psi_\alpha(\br + L\,\bd{e}_a)$ into twisted ones,
\bealn
U(\bkappa)^\dagger\,\Psi_\alpha(\br)\,U(\bkappa) =   
\esp{-i\bkappa\cdot\br}\;\Psi_\alpha(\br)\,. 
\eal
One finds that 
\beal
W(\bkappa) &= i\,\int d\br\, \bkappa\cdot\br\,\rho(\br)\,,
\label{W-1}
\eal
and translates by $\bkappa$ the total momentum of the many-body system. 
As emphasised by Kohn \cite{Kohn-1964}, despite after the gauge transformation 
$H(\bkappa)\to H$, its spectrum does depend on $\bkappa$ since it must be 
calculated on a basis of many-body wavefunctions satisfying  
\bealn
&\Psi(\br_1,\br_2,\dots,\br_i + L\,\bd{e}_a,\dots\br_N)\\
&\qquad = \esp{i\phi_a}\;\Psi(\br_1,\br_2,\dots,\br_i,\dots\br_N)\,,
\eal
which have not anymore the periodicity of the torus, unlike $H$ does. \\
In the case of an Hamiltonian with a HK interaction, as in \eqn{Ham}, 
there is a further issue. Indeed, while the non-interacting Hamiltonian 
$H_0$ has a first quantization representation, 
\bealn
H_0 &= \sum_{i=1}^N\,\bigg(\fract{\bp_i^2}{2m} + V(\br_i)\bigg)\,,
\eal
where $i=1,\dots,N$ label the $N$ electrons, so that $H_0(\bkappa)$ is simply obtained by $\bp_i\to \bp_i+\bkappa$, 
$U_\text{HK}$ does not and therefore it is not obvious what $U_\text{HK}(\bkappa)$ is. For that, one can reasonably assume, as we do hereafter, that the 
previous gauge argument can be extended even in HK Hamiltonians, thus that $U^\dagger(\bkappa)\,U_\text{HK}(\bkappa)\,U(\bkappa)= U_\text{HK}$, 
which implies that  
\beal
&U_\text{HK}(\bkappa) = U(\bkappa)\,U_\text{HK}\,U^\dagger(\bkappa)\\
&\; = \sum_{\alpha\beta\gamma\delta}\,\fract{1}{L^3}\int \prod_{i=1}^4\,d\br_i\;
U_{\alpha\beta\gamma\delta}(\br_1+\br_2-\br_3-\br_4)\\
&\; \esp{-i\bkappa\cdot(\br_1+\br_2-\br_3-\br_4)}\;
\Psi^\dagger_\alpha(\br_1)\,
\Psi^\dagger_\beta(\br_2)\,
\Psi^\dagga_\gamma(\br_3)\,
\Psi^\dagga_\alpha(\br_4)\,.\label{U(kappa)-1}
\eal
We observe that, unlike conventional two-body interaction terms, 
\eqn{U(kappa)-1} explicitly depends on $\bkappa$. In particular, $U_\text{HK}(\bkappa)$ looses the nice property of $U_\text{HK}$ of being diagonal in momentum.   
Indeed, is we write, consistently with the torus geometry,
\bealn
U_{\alpha\beta\gamma\delta}(\br) &= \fract{1}{L^3}\,\sum_\bq\,
U_{\alpha\beta\gamma\delta}(\bq)\,\esp{i\bq\cdot\br}\;,\\
\Psi^\dagga_\alpha(\br) &= \fract{1}{\;\sqrt{L^3\,}\;}\,\sum_\bk\, 
\esp{i\bk\cdot\br}\;c^\dagga_{\alpha\bk}\,,
\eal
where $\bq$ and $\bk$ have components quantised in integer multiples 
of $2\pi/L$, then 
\beal
U_\text{HK}(\bkappa) &= \sum_{\bq\,\bk_i}
U_{\alpha\beta\gamma\delta}(\bq)\, 
c^\dagger_{\alpha\bk_1}\,
c^\dagger_{\beta\bk_2}\,
c^\dagga_{\gamma\bk_3}\,
c^\dagga_{\delta\bk_4}\\
&\qquad I(\bq,\bk_1)\,I(\bq,\bk_2)\,I(\bq,\bk_3)^*\,I(\bq,\bk_4)^*\,,
\label{U(kappa)-2}
\eal
where
\bealn
I(\bq,\bk) &= \fract{1}{L^3}\,\int d\br\,\esp{i(\bq-\bk-\bkappa)\cdot\br}\;,
\eal
is not equal to $\delta_{\bk,\bq}$ unless $\bkappa=0$ or in the thermodynamic 
limit $L\to\infty$. \\
To proceed with Kohn's argument, we need to evaluate the ground state energy 
$E_0(\bkappa)$ of $H(\bkappa)$ and calculate its curvature at $\bkappa=0$. 
We assume that the Peierls substitution is valid for the non-interacting Hamiltonian \eqn{Ham-0} \cite{Toschi-PRB2012,DanielePRL2020,Marco-PRB2021,Deveraux-PRB2021}, so that  
\be
H_0(\bk,\bkappa) =\sum_{\alpha,\beta=1\dots M} t^{\alpha\beta}(\bk+\bkappa)\, c^\dagger_{\alpha\bk}\, c^\dagga_{\beta\bk}\,,
\label{Ham-0-flux}
\ee
which can be readily expanded up to second order in $\bkappa$. We further need 
to expand \eqn{U(kappa)-2} to the same order, which is equivalent to 
\be
U_\text{HK}(\bkappa) \simeq  U_\text{HK}-\big[W(\bkappa),U_\text{HK}\big]
+\fract{1}{2}\Big[W(\bkappa),\big[W(\bkappa),U_\text{HK}\big]\Big]\,.
\label{U(kappa) second order} 
\ee
Since $\rho(\br)$ in \eqn{W-1} has the periodicity of the torus, it can be written as 
\bealn
\rho(\br) &= \fract{1}{L^3}\,\sum_\bq\,\esp{i\bq\cdot\br}\;\rho(\bq)\,,&
\bq &= \fract{2\pi}{L}\big(n_x,n_y,n_z\big)\,.
\eal
Therefore $W(\bkappa)$ in \eqn{W-1} is also equivalent to   
\beal
W(\bkappa) &= \sum_{a=x,y,z}\,\sum_{q_a\not=0}\,
\fract{\kappa_a}{q_a}\;\rho\left(q_a\,\bd{e}_a\right)\,.\label{W-2}
\eal
and can be used to evaluate \eqn{U(kappa) second order}. Equations
\eqn{Ham-0-flux} and \eqn{U(kappa) second order} allow calculating  
the ground state energy up to second order in $\bkappa$. \\

\noindent
As an example, we consider a simple one-band model, with nearest-neighbour hopping $-t$ and HK interaction $U_{\rm HK}(\bk)=U(\bk)\,(n_{\bk}-1)^2/2$, where 
$n_\bk = n_{\bk\up} + n_{\bk\down}$. For $U(\bk)>4dt>0$, the model describes a Mott insulator where each momentum state is occupied by a single electron. This state 
is degenerate, since each electron can have any spin. We decide to use the non-pure 
density matrix 
\bealn
\rho=\prod_\bk \,\fract{\;\ket{\bk,\up}\bra{ \bk,\up} + \ket{\bk,\down}\bra{ \bk,\down}\;}{2}\;,
\eal
to perform the calculations, a choice commonly adopted in the literature. 
Since each $\bk$ is singly-occupied, the ground state energy variation 
at second order in $\bkappa$ 
comes just from the interaction part \eqn{U(kappa) second order} and, assuming the flux finite only along $x$, reads 
\bw 
\be
 \Delta E_0(\bkappa)\simeq E_0(\bkappa) -E_0(\bd{0})=\sum_{q_x\not=0}\,\sum_\bk\,\fract{\phi_x^2}{\;2L^2 q_x^2\;}\;
\fract{\; \Big(U(\bk)+U(\bk+q_x\bd{e}_x)\Big)\,\big(\epsilon_{\bk}-\epsilon_{\bk+q_x\bd{e}_x}\big) \;}{\; U(\bk)+U(\bk+q_x\bd{e}_x) +2\big(\epsilon_{\bk}-\epsilon_{\bk+q_x\bd{e}_x}\big)\;} \,.
\label{single-band-DeltaE}
\ee
\ew
Few comments are in order. First, $\Delta E(\bkappa)\not=0$ despite the single-particle gap, and vanishes only for $U(\bk)\to\infty$.\\
Furthermore, the double sum, over $q_x$ and $\bk$, grows as $L^{d+1}$, which corresponds through  
Eq.~\eqn{Kohn-Drude} to a Drude weight scaling as $L$, evidently a consequence of the infinite range of the HK interaction. \\ 
Finally, one easily realises that $\Delta E(\bkappa) <0$, which formally corresponds to a negative Drude weight linear in the system size, thus to a singular (orbital) paramagnetic response. In fact, we can rigorously prove that $\Delta E(\bkappa) <0$ whenever the eigenstates of $H(\bk)$ are simultaneously eigenstates of $H_0(\bk)$ and of $U_\text{HK}(\bk)$, which indeed occurs in the 
single-band model we analysed. In more general cases when $\big[H_0(\bk),U_\text{HK}(\bk)\big]\not=0$, we cannot exclude that $\Delta E(\bkappa) >0$, thus a positive Drude weight still growing with the linear size of the system. However, irrespective of the sign of $\Delta E(\bkappa)$, we can definitely conclude that a model with HK interaction sustains a finite current at any finite flux, diamagnetic if $\Delta E(\bkappa)>0$ and 
paramagnetic otherwise, even if it has a gap in the single-particle spectrum. This, in light of Kohn's argument, does not come as a surprise since the interaction has infinite range in real space and therefore also the insulator remains sensitive to the boundary conditions, as recently also discussed in Ref.~\cite{skolimowski2024realspace}.

\noindent
We conclude noticing that the knowledge of the ground-state wavefunction 
$\ket{\Psi_0(\bkappa)}$ up to first order in $\bkappa$ allows the direct evaluation 
of the many-body Chern number $C$, for instance in $d=2$ \cite{Resta-PRL2005}
\bealn
C &= -\lim_{L\to\infty}\,\fract{2\pi}{L^2}\,
\Ima\,\bra{\partial_{\kappa_x}\Psi_0(\bkappa)}
\partial_{\kappa_y}\Psi_0(\bkappa)\rangle_{\big|\bkappa=\bd{0}}\,,
\eal
which avoids using Kubo formulas.

\section{Concluding remarks}\label{sec4}
Models with Hatsugai-Kohmoto interactions~\cite{HK_original} are often used to 
discuss non-perturbative phenomena and represent, in particular, a very elegant and a convenient analytic proxy to Mottness.  
In general, the latter is instead hard to access and exactly solvable only in specific limits. 
However, in view of the infinite range of the HK interaction, one must be very cautious in adapting results valid 
for realistic models to HK ones, and vice versa. Indeed, we have shown that   
the common way of calculating thermodynamic susceptibilities and transport coefficients through 
Kubo formulas is incorrect when the interaction is of Hatsugai-Kohmoto type.

Important considerations regard also the topological properties of HK-interacting insulators. On the one hand, it has been shown that for HK models, similarly to the case of Hubbard-Mott insulators \cite{PhysRevB.108.125115,Tremblay-preprint2023}, quantum spin Hall conductivities can correctly be non-quantized and hence deviate from the values of the (interacting) topological invariant \cite{Phillips-PRL2023,Qimiao-preprint2023}. On the other hand, the issues of transport originating from the infinite-range nature of the HK interaction discussed in this work should be taken into account. In particular, the unphysical (and infinite) value of the diagonal bulk conductivity notwithstanding the presence of a gap as well as the impossibility, due to the infinite range of the interaction, to introduce a notion of boundary are pathologies that in some cases may reduce the convenience of using HK models for studying interacting topology.

\begin{acknowledgments}
We thank L. Benfatto, S. Ciuchi, M. Reitner, N. Wagner and A. Toschi for useful discussions.
This research was supported in part by grant NSF PHY-1748958 to the Kavli Institute for Theoretical Physics (KITP).
G. S. acknowledges support from the Deutsche Forschungsgemeinschaft (DFG, German Research Foundation) – Project-ID 258499086 – SFB 1170 and through FOR 5249-449872909 (Project P5). 
The Flatiron Institute is a division of the Simons Foundation.    
\end{acknowledgments}

\bibliographystyle{apsrev4-2}
\bibliography{bib}
\end{document}